\documentclass{nature-preprint}
\usepackage{tcolorbox,color}
\usepackage{colortbl}
\usepackage{ulem}
\usepackage{amsmath}
\usepackage{amssymb}
\usepackage{graphicx}
\usepackage{upgreek}

\usepackage{caption,wrapfig}
\usepackage[hidelinks]{hyperref}
\usepackage{lipsum}
\usepackage[labelfont=bf]{caption}
\usepackage{color,soul}
\usepackage[super,comma,compress]{natbib}

\title{2D semiconductors as on-chip light sources for integrated nanophotonics}

\author{Christian Frydendahl$^{1,2,*}$, Torgom Yezekyan$^{1,3}$, Vladimir A. Zenin$^1$, and Sergey I. Bozhevolnyi$^{1,*}$}

\begin{document}

\maketitle 

\begin{affiliations}
\small
\item Centre for Nano Optics, University of Southern Denmark, Campusvej 55, DK-5230, Odense, Denmark
\item Department of Physics and Astronomy, Aarhus University, Ny Munkegade 120, DK-8000, Aarhus, Denmark
\item POLIMA - Center for Polariton-driven Light–Matter Interactions, University of Southern Denmark, Campusvej 55, DK-5230, Odense, Denmark
\item[]
\item[*] cfry@phys.au.dk and seib@mci.sdu.dk
\end{affiliations}


\begin{abstract}
\textbf{Abstract:} Incorporating on-chip light sources directly into nanophotonic waveguides generally requires introducing a different material to the chip than that used for guiding the light, a crucial step that requires dealing with several technical challenges, e.g., atomic lattice mismatch in epitaxial growth between substrate and luminescent materials resulting in strain defects that lower performance. Here we demonstrate that van der Waals materials, such as the 2D semiconductor MoSe$_2$, can easily be transferred onto gold plasmonic nanowaveguides using standard dry visco-elastic polymer transfer techniques. We further show that the photoluminescence from MoSe$_2$ can be injected directly into these on-chip waveguides. Our fabrication methods are compatible with large-scale roll-to-roll manufacturing techniques, highlighting a potential cost-effective and scalable hybrid plasmonic and 2D semiconductor platform for integrated nanophotonics.
\end{abstract}

\textbf{Keywords:} Plasmonics, 2D materials, integrated photonics, on-chip light sources, Purcell enhancement


Efficient optical connection between integrated optics and the outside world remains one of the biggest challenges for widespread commercialization and adoption of photonic integrated circuits (PICs)\cite{zhou:2023,yang:2023}. To inject light into integrated nanophotonic waveguides, it must be coupled from the outside with a lensed fiber or microscope objective. This limits the scalability and ultimate promise of independent optical chips. Making direct, on-chip, light sources that directly inject radiation into the waveguide circuitry would circumvent the problem of external connection, making a crucial step towards delivering on numerous promises of future PICs, e.g., faster and more energy efficient data processing and communications\cite{zhou:2023,yang:2023}, artificial neural networks\cite{cheng:2020,frydendahl:2021}, diagnostic lab-on-chip systems\cite{rodriguez:2016}, and new chip-scale quantum technologies\cite{siampour:2018,blauth:2018,wang:2020}.

The central problem for on-chip sources (and photodetectors) is that the waveguiding material should exhibit low absorption and emission at the working wavelength, a requirement that is at odds with realization of efficient light sources and detectors\cite{yang:2023,zhou:2023,yezekyan:2021}. Heterogeneous integration is by no means an easy solution. For example, one can exploit epitaxial growth of III-V semiconductors onto the photonic waveguide substrate, but realizing high-quality crystals with efficient luminescence is cumbersome due to strain and stress related defects in such devices\cite{yang:2023,zhou:2023,liu:2018}. One can utilize instead wafer bonding to integrate the III-V materials into the optical circuitry, but this approach comes at the price of reduced scalability\cite{yang:2023}. Another method proposed is to locally dope the waveguiding material, thereby locally changing material properties and introducing emissive/absorptive defects\cite{rodriguez:2014,gherabli:2020}. These devices however generally suffer from low quantum efficiencies and poor performance metrics. A promising strategy, which is also seeing commercialization, is that of chiplets: optical circuits assembled from groups of individual smaller chips made with different processes and materials to maintain (close to) ideal device performance\cite{fotouhi:2019,huang:2024}. While a promising strategy from the performance perspective, this approach comes with the price of new difficulties for large-scale fabrication and cost-effectiveness\cite{fotouhi:2019,huang:2024}.

One candidate however has attracted surprisingly little attention so far: 2D transition metal dichalcogenides (TMDCs). 2D materials are compatible with most material substrates as they adhere to other materials strictly by van der Waals interactions\cite{gherabli:2023,liu:2016}, i.e., there is no need for chemical compatibility or any lattice constant matching as in conventional heterostructure devices. This, together with the large variety of band gaps found in TMDCs\cite{Rasmussen:2015} makes them uniquely interesting from the viewpoint of realization of on-chip light sources, as several material platforms (gold, silicon, silicon nitride, indium phosphide, etc.) are being developed for the PIC industry depending on the application and optical wavelength\cite{huang:2024}. 2D materials, with their substrate agnostic nature, could thus provide a universal on-chip light source (and detector) approach for several upcoming PIC platforms. Notably, 2D material heterostructures have already exhibited excellent optoelectronic device performance metrics, approaching or superseding commercial technologies\cite{gherabli:2023}. While there has already been several investigations into integrating 2D materials with nanophotonic waveguides for optical modulation\cite{liu:2011,phare:2015}, on-chip detection\cite{gherabli:2023,flory:2020,marin:2019}, or even utilizing such materials themselves for waveguiding\cite{ling:2023,li:2021,dolado:2020}, there is still a lack of examples in the literature towards utilizing their luminescence for waveguiding in integrated nanophotonics

Recently, there has been several promising advances towards wafer-scale production of single crystal TMDC monolayers using chemical vapor deposition\cite{kim:2023,li:2024} as well as growth of advanced heterostructures and junction devices such as light emitting diodes (LEDs)\cite{najafidehaghani:2021}. These developments along with recent advances in roll-to-roll fabrication and automated 2D material transfer systems\cite{cai:2018,masubuchi:2018,mannix:2022} point towards promising technological applications of 2D materials in the near future\cite{kong:2019,kim:2023}.

However, particular care must be taken when trying to utilize TMDC monolayers as on-chip light sources. For one, the TMDC layer thickness is extremely small, implying that waveguiding configurations with small optical mode volumes should be utilized in order to maximize optical coupling. Secondly, the in-plane confinement of the excitons that dominate the TMDC optical emission restricts the optical coupling to transverse electric (TE) optical modes. Considering various nanophotonic waveguiding configurations, plasmonic slot waveguides\cite{andryieuski:2012,smith:2015,ayata:2017} appear to be one of the most suitable configurations for efficient coupling of emission from TMDC monolayers.

Surface plasmon polaritons (SPPs), often shortened to surface plasmons, represent collective oscillations of conduction electrons in metals coupled to electromagnetic fields in neighboring dielectrics that are bound to and propagate along metal/dielectric interfaces. Plasmonic slot waveguides enable SPP guiding by confining SPP fields within metallic channels and offer control of guided modes similar to that exercised in dielectric waveguides, e.g., optical fibers/silicon nanophotonic waveguides, etc. However, unlike dielectric waveguides, plasmonic waveguides do not rely on total internal reflection and can facilitate mode sizes below the diffraction limit\cite{smith:2015}. Thus, these waveguides can be scaled down to few nanometer-sized cross sections, resulting in extremely confined optical fields. This extreme optical confinement results in strongly enhanced electromagnetic interactions with emitters that are physically small, such as molecules\cite{kumar:2020}, quantum dots\cite{siampour:2018}, and 2D materials\cite{blauth:2018,gherabli:2023}. This extreme miniaturization is also advantageous for optoelectronics, since smaller components result in smaller device capacitances, thus lowering electrical RC-time constants which increases the possible operation bandwidths\cite{ayata:2017,thomaschewski:2021,yezekyan:2024}.

In this study, we demonstrate the efficient coupling and guiding of photoluminescence from monolayers of MoSe$_2$ into gold plasmonic slot waveguides. In the investigated 'proof of concept' configuration, optical emission from the 2D material is stimulated by far-field optical pumping with a 620\,nm laser, while our results are applicable to emission via electroluminescence as well. By attaching a plasmonic antenna coupler that is resonant with the emission wavelength of MoSe$_2$, we make use of the Purcell effect to enhance the coupling of optical emission to the plasmonic slot waveguide's fundamental mode. We confirm this coupling to the fundamental mode via spectral and polarization sensitive measurements. Our results highlight the mutual compatibility between plasmonic waveguide systems and 2D semiconductors, indicating a promising future direction for fully independent PIC systems based on 2D material light-sources and plasmonic waveguides.

\begin{figure}[h]
    \centering
    \includegraphics[width=1\linewidth]{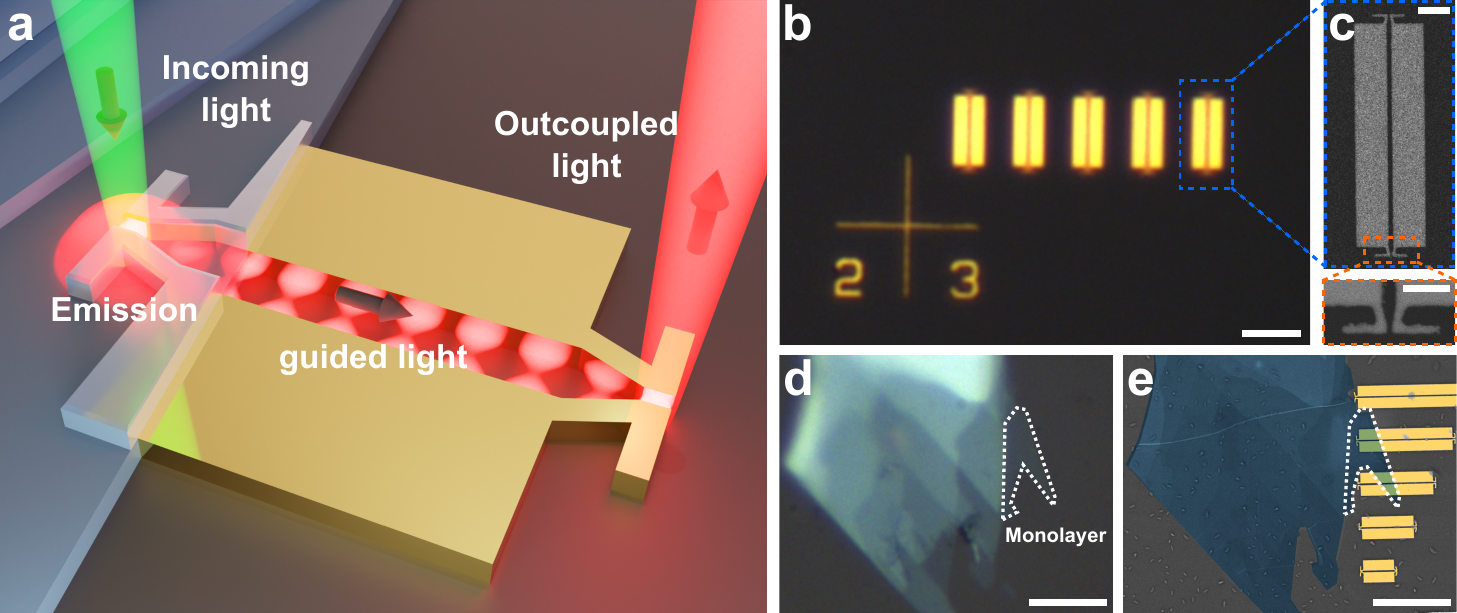}
    \caption{\textbf{Experimental configuration.} \textbf{a)} Schematic of the experiment. Shorter wavelength light is used to stimulate photoluminescence from MoSe$_2$ attached to a plasmonic waveguide. Part of the emitted light goes into the guided mode of the waveguide, and is out-coupled at the terminal dipole antenna to free-space. \textbf{b)} Optical mircrograph of fabricated plasmonic waveguides. Scale bar is 5\,µm. \textbf{c)} SEM image of a waveguide, with inset showing the coupling antenna and tapered section. Scale bar is 1\,µm, and in inset scale bar is 500\,nm. \textbf{d)} Optical image of an exfoliated layer of MoSe$_2$, with monolayer highlighted. Scale bar is 5\,µm. \textbf{e)} False color SEM image of the same flake, transferred to a set of waveguides. The pill-shaped defects are contaminants accumulated after removal of the encapsulating PC film. Scale bar 5\,µm.}
    \label{fig1}
\end{figure}

Figure~\ref{fig1}.a shows a schematic of the experimental configuration: a plasmonic slot waveguide is defined by two slabs of gold on a glass substrate that have a gap between them of 150\,nm. At each end of the waveguide, tapered sections connect the waveguide structure to plasmonic gap antennas\cite{andryieuski:2012} with a gap width of 50\,nm. At one set of antennas, a sheet of monolayer MoSe$_2$ has been transferred on top (left side). Light with a wavelength shorter than the band gap of MoSe$_2$ is then illuminated onto the antenna/MoSe$_2$ section. The light is absorbed, and photoluminescence is emitted from radiative recombination of the optically generated excitons and trions formed by the generated electron-hole pairs in the MoSe$_2$. Part of this emission will directly emit as the fundamental mode of the slot waveguide. It will then propagate towards the opposite pair of gap antennas (right side of Fig.~\ref{fig1}.a), where it will be radiated out to the far-field for an external observer to record it.

Examples of fabricated waveguides can be seen in Fig.~\ref{fig1}.b and c (see more details in Methods). After the waveguides have been fabricated, exfoliation and transfer of MoSe$_2$ monolayers is performed with a standard visco-elastic transfer method using a sacrificial layer of polycarbonate (PC), Fig.~\ref{fig1}.d and e (also detailed in Methods below). This puts the MoSe$_2$ into contact with the coupling antennas from above, ensuring optical emission enhancement\cite{kewes:2016}, see also Supplementary Fig.~1 and surrounding discussion. The transfers are performed in such a way that the MoSe$_2$ monolayers only cover part of the waveguides: this is done to minimize the propagation losses associated with reabsorption into the MoSe$_2$. The highly confined mode of the slot waveguide is concentrating the optical energy such that it is easily reabsorbed into the MoSe$_2$ layer, and thus too much physical overlap of the MoSe$_2$ layer and the waveguide should generally be avoided. One thing to note is that the transfer method naturally results in an encapsulated device, as we transfer by melting the PC layer the flake is carried by directly onto the waveguide substrate, leaving a thin PC film. After melting the PC off to transfer the MoSe$_2$ we get a sample where we have matched the refractive index above and below the gold nanowaveguides (as PC's refractive index of, $n_\text{PC}\sim1.58$, is closely matched to glass: $n_\text{glass}\sim1.5$), this ensures that the optical mode stays confined in the metal gap. This also has the added benefit of protecting the sample from degradation (MoSe$_2$ oxidation) in ambient conditions, as indeed we observed samples that perform to similar levels even after several months of fabrication.

\begin{figure}[h]
    \centering
    \includegraphics[width=0.95\linewidth]{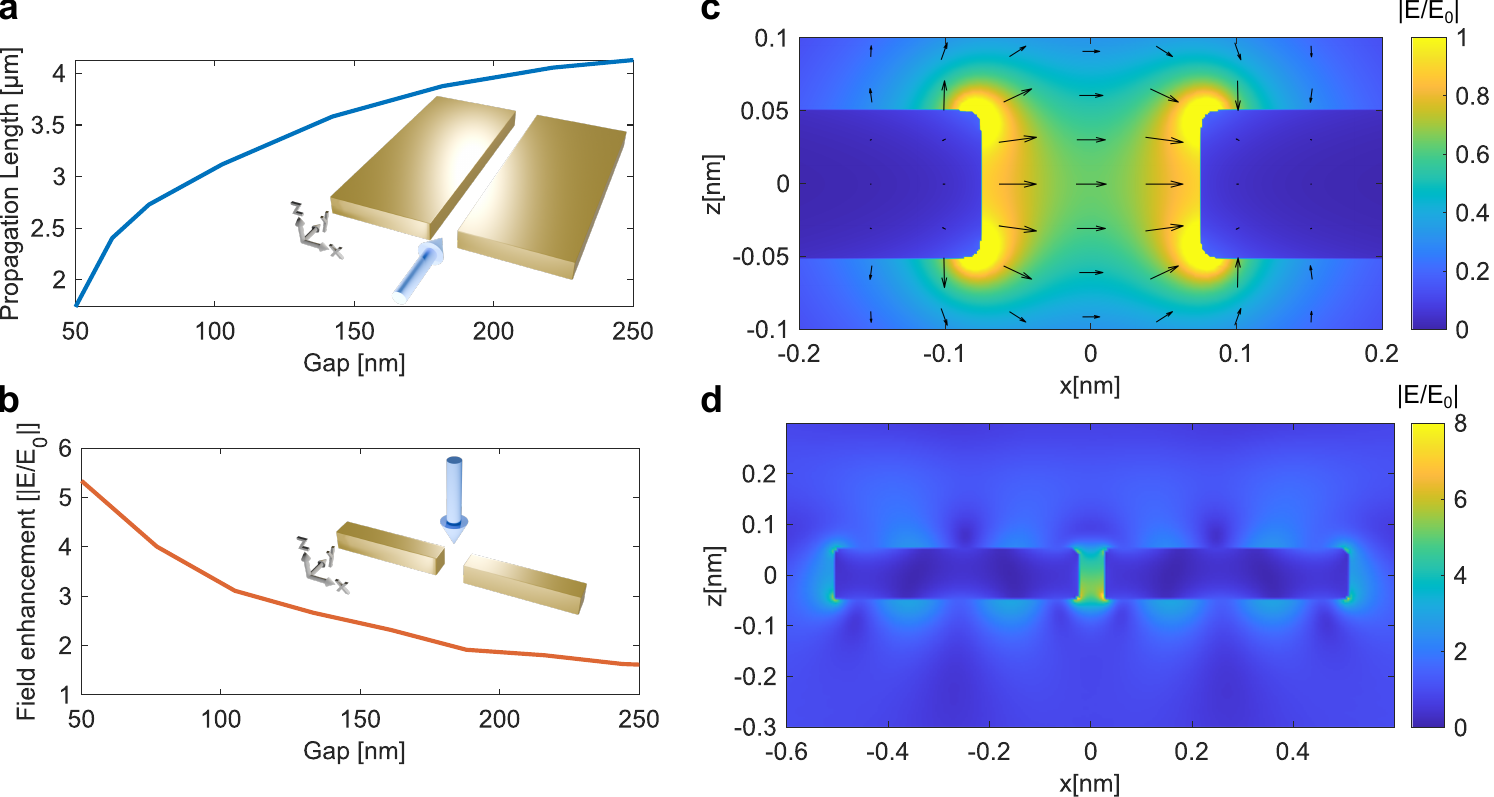}
    \caption{\textbf{Plasmonic waveguide design.} \textbf{a)} Waveguide propagation length and \textbf{b)} field enhancement factor of gap antenna, both for varying gap sizes. The inverse relationship between the two shows why a tapered coupling section is needed to maximize optical in-coupling and maintain longer transmission. The blue arrows in the insets indicate the optical injection axis in the simulations.} \textbf{c)} Simulations of the electric field distribution of the fundamental mode of a plasmonic slot waveguide. Arrows indicate electric field direction, showing the field is strongly aligned with the $x$-axis inside the gap. \textbf{d)} Simulation of the near-field of a gap antenna under plane wave illumination, showing optical field enhancement in the gap.
    \label{fig2}
\end{figure}

The coupling antennas serve two purposes in our structure: 1) they match in-coming (and out-going) far-field radiation to the fundamental mode of the waveguide, and 2) they enhance the total emission from the MoSe$_2$ monolayer via Purcell enhancement. After absorption of a photon, the created electron-hole pair in the MoSe$_2$ can decay along multiple channels, each with an associated decay rate\cite{kumar:2021,kewes:2016}: radiation to the far-field, $\Gamma_\text{rad}$, non-radiative recombination, $\Gamma_\text{nr}$, or emission into the antenna gap mode, $\Gamma_\text{pl}$. A portion of the radiation in the gap mode will then be transmitted to the waveguided mode\cite{andryieuski:2012} - maximizing the Purcell enhancement of the gap antenna will thus increase the overall optical coupling to the guided mode.

However, it is well-known that maximal field enhancement is seen for small antenna gaps, while maximal propagation in plasmonic slot waveguides is seen for wider gaps. Fig.~\ref{fig2}.a and b shows respectively the propagation length in a slot waveguide and field enhancement in a gap antenna for various gap widths, both for excitation at 800\,nm (matching the emission of MoSe$_2$). We have picked an antenna gap of 50\,nm and a waveguide gap of 150\,nm for our device (the electric field profiles can be seen in Fig.~\ref{fig2}.c and d), and utilize a 200\,nm long tapered section to smoothly convert between the two with minimal mode conversion loss due to their similarity in mode profiles (see Supplementary Fig.~2). It should however be noted that our experimental devices are likely to have lower overall propagation and field enhancement, due to surface roughness effects in the fabrication\cite{huang:2010} that are unaccounted for in the simulations. It should be emphasized: the emission of the MoSe$_2$ is enhanced directly into the guided mode of the waveguides by the coupling antennas, which is expected to ensure very efficient coupling\cite{kumar:2021,chen:2010,kewes:2016}. Indeed, coupling efficiencies close to 100\% have been reported for emitters directly inside plasmonic waveguides before\cite{kewes:2016,kumar:2021}.

After transferring the MoSe$_2$ to the fabricated waveguides we generally see a bright hot-spot in photoluminescence images originating at the center of the nanoantennas (see Supplementary Fig.~3) which is a strong indication that the MoSe$_2$ is well-coupled to the antennas.

\begin{figure}[h]
    \centering
    \includegraphics[width=0.95\linewidth]{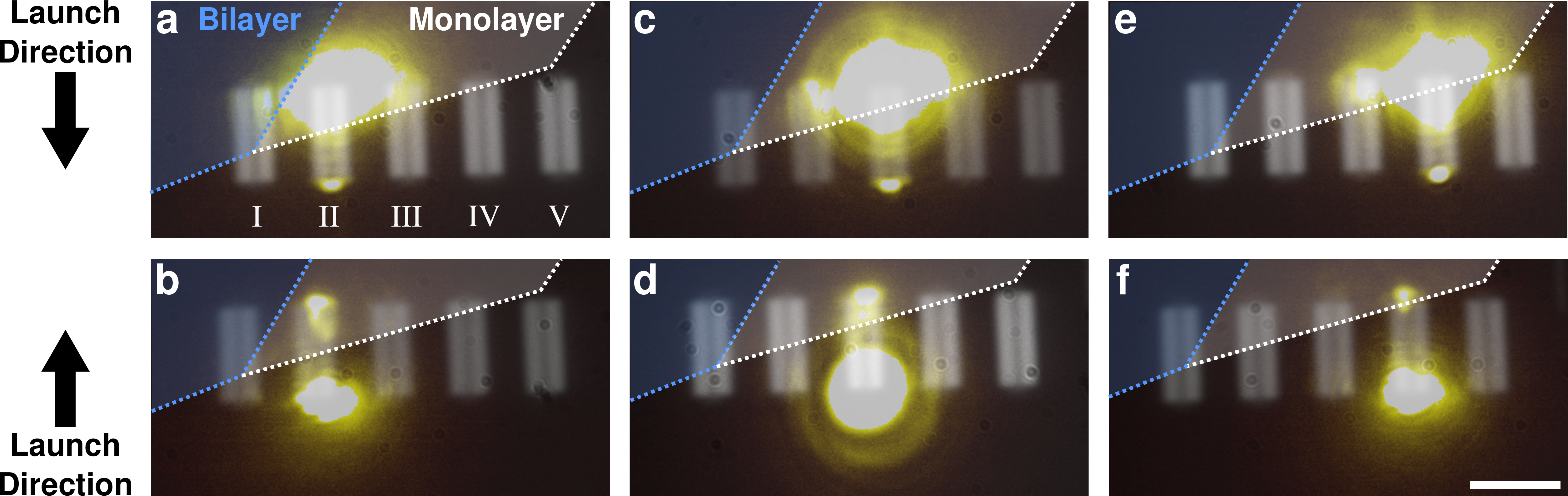}
    \caption{\textbf{Optical propagation measurements.} \textbf{a)}, \textbf{c)}, and \textbf{e)}, shows the out-coupling of radiation from the opposite dipole antennas when the MoSe$_2$-monolayer coupled to the waveguides is optically excited. \textbf{b)}, \textbf{d)}, and \textbf{f)} shows radiation coupling out from the dipole antennas coupled to the MoSe$_2$-layer when light is injected from the opposite end of the waveguide. For both cases, a long-pass filter is inserted in front of the camera to block the majority of the excitation laser. The 5 waveguides are labeled I-V from left to right for general refence. Scale bar is 5\,$\upmu$m. Images are composites of brightfield images of the waveguides and the PL images with only laser excitation.}
    \label{fig3}
\end{figure}

Having transferred MoSe$_2$ to the waveguides, we perform optical propagation measurements, see Fig.~\ref{fig3} (the full optical system is described in Methods and Supplementary Fig.~4). We investigate the injection of light from both ends of the waveguides. Fig.~\ref{fig3}.a, c, e show the case for illuminating the system from the MoSe$_2$ side and collecting the out-coupled photoluminescence, and Fig.~\ref{fig3}.b, d, and f show the case for injecting light at the bare antenna side and seeing the scattered photoluminescence and antenna-coupled emission from the MoSe$_2$. Despite putting a long-pass filter to block the excitation laser from the camera, for both cases we see a reflection from the excitation laser, with a smaller spot corresponding to the out-coupled signal from the antennas at the opposite end of the waveguides (an example of laser leakage spectral rangeis shown in Supplementary Fig~5). We also see that the out-coupled signal is generally strongest for waveguide IV (in either launch direction), where the smallest amount of MoSe$_2$ is touching the waveguide - in agreement with our previous assessment that MoSe$_2$ inside the waveguide section will result in larger reabsorption losses (see Supplementary Fig.~6 for more details). Notably, we also see the coupling of photoluminescence from the MoSe$_2$ bi-layer region in the sample at waveguide I, albeit at a weaker overall intensity than for the monolayer case (Supplementary Fig~7).

\begin{figure}[h]
    \centering
    \includegraphics[width=1\linewidth]{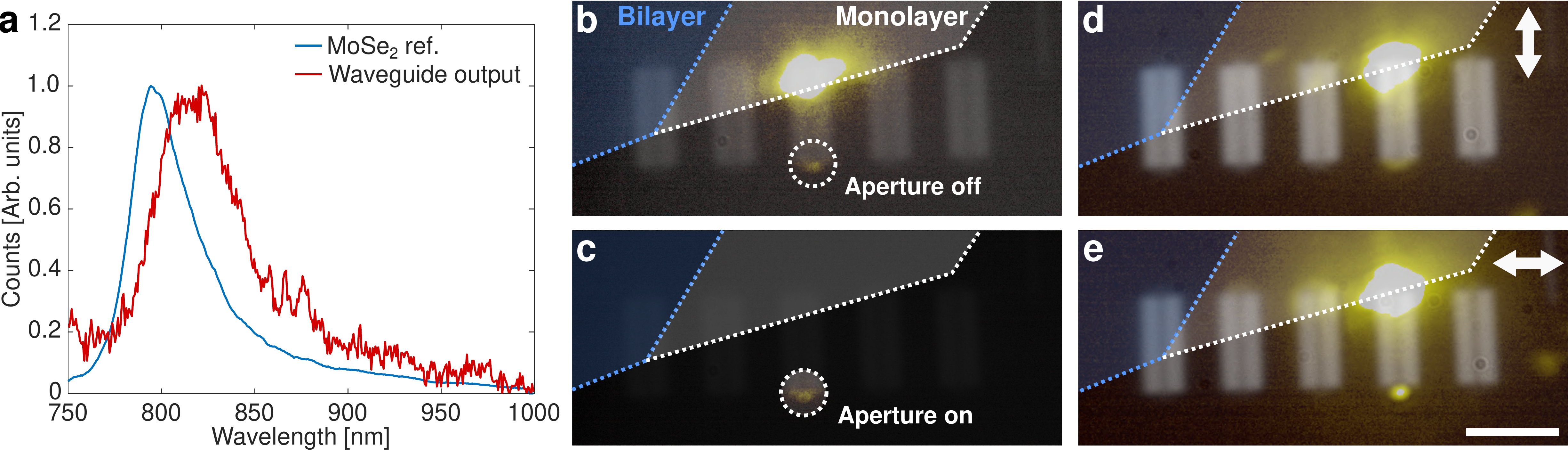}
    \caption{\textbf{Spectral composition and polarization of outcoupled light.} \textbf{a)} Reference spectrum of the same MoSe$_2$-flake and the spectrum recorded at the output of the waveguide in \textbf{b)} and \textbf{c)} when a spatial aperture is inserted into the system to cut the light going to the spectrometer such that only the outcoupled emission is measured. \textbf{d)} and \textbf{e)} show images recorded with a linear polarization filter in front of the camera respectively at cross- and parallel orientation to the optical polarization of the outcoupling dipole antenna. Scale bar is 5\,$\upmu$m. Images are composites of brightfield images of the waveguides and the PL images with only laser excitation.}
    \label{fig4}
\end{figure}

To confirm that the out-coupled signal is indeed the photoluminescence of the MoSe$_2$ and if it is coupled to the waveguide's fundamental mode (or merely scattered) we perform spectral- and polarization measurements. To record the optical spectrum coming from the sample we utilize a flip-mirror to direct the sample image to a fiber-coupled spectrometer in the same optical system (Supplementary Fig. 4).

Fig.~\ref{fig4}.a shows a reference spectrum from the same MoSe$_2$-flake from a different area of the sample without any waveguides or nanoantennas (blue line). The peak exhibits the characteristic asymmetric double Lorentzian line-shape of mixed exciton and trion emission from MoSe$_2$\cite{chen:2016,indukuri:2020}. Next, we insert an aperture into the optical path to perform spatial filtering (see Supplementary Fig.~4) such that we can remove all but the light from the out-coupling antennas going to the spectrometer. In this way, we can block any potentially scattered photoluminscence from the MoSe$_2$ monolayer, and isolate only the out-coupled signal. The area encompassed by the aperture can be seen in Fig.~\ref{fig4}.b and c, and the resulting spectrum can be seen in Fig.~\ref{fig4}.a (red line). 

We see generally the spectral shape of the out-coupled signal is reminiscent of the reference, however it has been red-shifted. One explanation for this is that the absorption losses in the waveguide are strongly dispersive: gold absorbs significantly more for shorter wavelengths than longer wavelengths. When the out-coupled signal is normalized to the same magnitude as the reference, the result is then an overall red-shifted appearance. See Supplementary Fig.~8 and surrounding discussion for more details.

The final test we perform is to verify the polarization of the observed bright spot at the out-coupling antennas. The fundamental mode of the waveguide is strongly polarized with its electrical field parallel to the substrate, and the nanoantenna couplers are also matched to this polarization to ensure good optical coupling between the antenna gap plasmon resonance and the waveguide mode. As such, if the out-coupled photoluminescence that we see is strongly polarized to match this polarization, we can confirm the light is coupled to the fundamental slot waveguide mode, and not just merely scattered across the sample (or coupled to some other guided SPP-mode hosted on the gold structures). Fig.~\ref{fig4}.d and e shows the results of inserting a linear polarizer in front the optical path before the camera, with Fig.~\ref{fig4}.d being cross-polarized to the fundamental mode and antenna resonance, while Fig.~\ref{fig4}.e shows the case for parallel polarization. We see clearly that the bright spot is suppressed for cross-polarization, while we see a clear and well-defined Gaussian spot (see Supplementary Fig.~9) for the parallel polarization, confirming that the light we see is indeed very strongly coupled to the guided mode.


In this work we have opted to use plasmonic gap antennas for the optical in-coupling of MoSe$_2$ photoluminescence. This has largely been done from practical considerations. 1) a smaller device footprint means we can attach the MoSe$_2$ monolayer to multiple waveguides in a single transfer, thus letting us test multiple devices within one fabrication, and 2) as we also need efficient out-coupling in our experiment to measure the transmitted signal, it is simplest to use a symmetric device with the same coupling geometry at each end. For far-field radiation, the overall coupling efficiency for such antennas is expected to be around 10\%\cite{andryieuski:2012}, which, assuming reciprocity, means roughly 10\% of transmitted MoSe$_2$ PL is out-coupled. The amount of in-coupled radiation is however more difficult to address, as the coupling antennas will enhance emission both to the far-field and to the guided mode. More detailed numerical studies are required to understand this aspect. However, what we observe is that there is a degree of coupling of photoluminescence directly from the MoSe$_2$ inside the waveguide sections themselves, which implies something akin to a candelabra\cite{volkov:2009} structure may be (albeit at a larger footprint) a more optimal in-coupling design. Near-field coupling of emitters directly to guided plasmonic modes have also shown close to 100\% coupling efficiency before\cite{kumar:2021,kewes:2016}.

We have demonstrated the mutual compatibility of plasmonic slot waveguides and MoSe$_2$ monolayers, and our results would also apply in general to other TMDCs. The fundamental mode of the waveguides is perfectly polarized to accept the emission from exciton and trion radiative recombination in the monolayer, as the bright exciton in-plane dipole moment is oriented to be in the plane of the TMDC-flake. While we have utilized photoluminescence here, our results can be generalized to, e.g., electroluminescence from a TMDC PN-junction, i.e., a light emitting diode or even a TMDC-based laser.


In this work we have demonstrated that the emission from MoSe$_2$ monolayers that are externally pumped with a laser can be coupled directly, on-chip, into plasmonic slot waveguides. Although more detailed investigations are desirable, our results indicate that TMDC monolayers can be found suitable for future on-chip light sources directly integrated with PICs by using plasmonic interfacing waveguides, which efficiently extract light from TMDCs and deliver it further to dielectric nanophotonic circuitry\cite{kewes:2016}. The fact that we see noticeably increased propagation losses due to waveguide sections being covered with MoSe$_2$ indicates that TMDCs can also be utilized in the same scheme for on-chip light detection with properly designed electrode configurations. Overall, given the compatibility of our fabrication methods with large-scale roll-to-roll manufacturing techniques, the considered hybrid plasmonic and TMDC-based platform appears to represent a potentially attractive cost-effective and scalable approach to integrated nanophotonics.


\section*{Methods}

\textbf{Plasmonic waveguide fabrication:} Plasmonic waveguides were fabricated using EBL. First, the borosilicate glass substrates were cleaned with ultrasonication in acetone, followed by an IPA and water rinse. Then, the substrates were covered with a $\sim$350\,nm PMMA resist, and EBL exposure was done. After development, 2\,nm Ti followed by 100\,nm Au was evaporated followed by lift-off of the remaining PMMA in acetone overnight. The waveguides were designed to have a 150\,nm gap in the slot, and a 50\,nm gap between the antennas.

\textbf{SEM images:} To record SEM images of the waveguides, the encapsulating PC film was first removed by soaking the sample in chloroform for $\sim$15 minutes. After this, a conductive polymer layer was spin-coated onto the sample to prevent charging effects during SEM imaging.

\textbf{FDTD simulations:} Optical simulations were performed using Lumerical FDTD. Propagation constants of plasmonic waveguides were found using the Finite Difference Eigenmode solver to find the fundamental mode of a waveguide cross section with varying gaps and a 100\,nm Au thickness for the metal layer, with a uniform surrounding dielectric of $n=1.5$ for the rest of the simulation. For nanoantenna simulations, a gap antenna with 50\,nm width and 480\,nm length was used, with a height of 100\,nm and a varying gap. Boundary conditions were Perfectly Matched Layers on all sides. The structure was illuminated with a plane wave from above with the electrical field polarized along the length of the antennas. Surrounding medium was a uniform dielectric with $n=1.5$, and gold's optical properties were taken from ref \citeonline{johnson:1972}.
  
\textbf{2D material transfer and exfoliation:} 2D material transfer to the waveguides was done using a standard dry visco-elastic transfer method in a custom-built 2D material transfer stage. See Supplementary Fig.~10 and surrounding discussion for more details. 

\textbf{Optical measurements:} Optical measurements were performed using a custom reflection microscope system, see Supplementary Fig.~4 and surrounding discussion for more details. All measurements were done in ambient atmosphere and at room temperature.

\vspace{12pt}

\textbf{Supporting Information:} Purcell enhancement simulations, high resolution mode field profiles, exfoliation and transfer information, schematic of measurement setup, laser leakage spectrum, propagation characteristics, guided bi-layer luminescence, propagation loss dispersion, Gaussian emission shape, 2D material transfer process schematic.

\bibliography{main}

\begin{thebibliography}{10}

\bibitem{zhou:2023}
Zhican Zhou, Xiangpeng Ou, Yuetong Fang, Emad Alkhazraji, Renjing Xu, Yating Wan, and John~E Bowers.
\newblock Prospects and applications of on-chip lasers.
\newblock {\em eLight}, 3(1):1, 2023.

\bibitem{yang:2023}
Junjie Yang, Mingchu Tang, Siming Chen, and Huiyun Liu.
\newblock From past to future: on-chip laser sources for photonic integrated circuits.
\newblock {\em Light Sci. Appl.}, 12(1):16, 2023.

\bibitem{cheng:2020}
Qixiang Cheng, Jihye Kwon, Madeleine Glick, Meisam Bahadori, Luca~P Carloni, and Keren Bergman.
\newblock Silicon photonics codesign for deep learning.
\newblock {\em Proc. IEEE}, 108(8):1261--1282, 2020.

\bibitem{frydendahl:2021}
Christian Frydendahl, SRK~Chaitanya Indukuri, Meir Grajower, Noa Mazurski, Joseph Shappir, and Uriel Levy.
\newblock Graphene photo memtransistor based on cmos flash memory technology with neuromorphic applications.
\newblock {\em ACS Photonics}, 8(9):2659--2665, 2021.

\bibitem{rodriguez:2016}
Isaac Rodríguez-Ruiz, Tobias~N. Ackermann, Xavier Mu{\~n}oz-Berbel, and Andreu Llobera.
\newblock Photonic lab-on-a-chip: Integration of optical spectroscopy in microfluidic systems.
\newblock {\em Anal. Chem.}, 88(13):6630--6637, 2016.

\bibitem{siampour:2018}
Hamidreza Siampour, Shailesh Kumar, Valery~A Davydov, Liudmila~F Kulikova, Viatcheslav~N Agafonov, and Sergey~I Bozhevolnyi.
\newblock On-chip excitation of single germanium vacancies in nanodiamonds embedded in plasmonic waveguides.
\newblock {\em Light Sci. Appl.}, 7(1):61, 2018.

\bibitem{blauth:2018}
M{\"a}x Blauth, M~J\"urgensen, Gwena{\"e}lle Vest, Oliver Hartwig, Maximilian Prechtl, John Cerne, Jonathan~J Finley, and Michael Kaniber.
\newblock Coupling single photons from discrete quantum emitters in wse2 to lithographically defined plasmonic slot waveguides.
\newblock {\em Nano Lett.}, 18(11):6812--6819, 2018.

\bibitem{wang:2020}
Jianwei Wang, Fabio Sciarrino, Anthony Laing, and Mark~G Thompson.
\newblock Integrated photonic quantum technologies.
\newblock {\em Nat. Photonics}, 14(5):273--284, 2020.

\bibitem{yezekyan:2021}
Torgom Yezekyan, Martin Thomaschewski, and Sergey~I Bozhevolnyi.
\newblock On-chip ge photodetector efficiency enhancement by local laser-induced crystallization.
\newblock {\em Nano Lett.}, 21(18):7472--7478, 2021.

\bibitem{liu:2018}
Alan~Y Liu and John Bowers.
\newblock Photonic integration with epitaxial iii--v on silicon.
\newblock {\em IEEE J. Sel. Top. Quantum Electron.}, 24(6):1--12, 2018.

\bibitem{rodriguez:2014}
Jos{\'e}~Antonio Rodr{\'\i}guez, Marco~Antonio V{\'a}squez-Agust{\'\i}n, Alfredo Morales-S{\'a}nchez, and Mariano Aceves-Mijares.
\newblock Emission mechanisms of si nanocrystals and defects in sio2 materials.
\newblock {\em J. Nanomater.}, 2014(1):409482, 2014.

\bibitem{gherabli:2020}
Rivka Gherabli, Meir Grajower, Joseph Shappir, Noa Mazurski, Menachem Wofsy, Naor Inbar, Jacob~B Khurgin, and Uriel Levy.
\newblock Role of surface passivation in integrated sub-bandgap silicon photodetection.
\newblock {\em Opt. Lett.}, 45(7):2128--2131, 2020.

\bibitem{fotouhi:2019}
Pouya Fotouhi, Sebastian Werner, Jason Lowe-Power, and SJ~Ben Yoo.
\newblock Enabling scalable chiplet-based uniform memory architectures with silicon photonics.
\newblock In {\em Proc. Int. Symp. Mem. Syst.}, pages 222--334, 2019.

\bibitem{huang:2024}
Huiyu Huang, Zhitian Shi, Giuseppe Talli, Maxim Kuschnerov, Richard Penty, and Qixiang Cheng.
\newblock Photonic chiplet interconnection via 3d-nanoprinted interposer.
\newblock {\em Light: Adv. Manuf.}, 5(LAM2024010011):1, 2024.

\bibitem{gherabli:2023}
Rivka Gherabli, SRKC Indukuri, Roy Zektzer, Christian Frydendahl, and Uriel Levy.
\newblock Mose2/ws2 heterojunction photodiode integrated with a silicon nitride waveguide for near infrared light detection with high responsivity.
\newblock {\em Light Sci. Appl.}, 12(1):60, 2023.

\bibitem{liu:2016}
Yuan Liu, Nathan~O Weiss, Xidong Duan, Hung-Chieh Cheng, Yu~Huang, and Xiangfeng Duan.
\newblock Van der waals heterostructures and devices.
\newblock {\em Nat. Rev. Mater.}, 1(9):1--17, 2016.

\bibitem{Rasmussen:2015}
Filip~A Rasmussen and Kristian~S Thygesen.
\newblock Computational 2d materials database: electronic structure of transition-metal dichalcogenides and oxides.
\newblock {\em J. Phys. Chem. C}, 119(23):13169--13183, 2015.

\bibitem{liu:2011}
Ming Liu, Xiaobo Yin, Erick Ulin-Avila, Baisong Geng, Thomas Zentgraf, Long Ju, Feng Wang, and Xiang Zhang.
\newblock A graphene-based broadband optical modulator.
\newblock {\em Nature}, 474(7349):64--67, 2011.

\bibitem{phare:2015}
Christopher~T Phare, Yoon-Ho Daniel~Lee, Jaime Cardenas, and Michal Lipson.
\newblock Graphene electro-optic modulator with 30 ghz bandwidth.
\newblock {\em Nat. Photonics}, 9(8):511--514, 2015.

\bibitem{flory:2020}
Nikolaus Fl{\"o}ry, Ping Ma, Yannick Salamin, Alexandros Emboras, Takashi Taniguchi, Kenji Watanabe, Juerg Leuthold, and Lukas Novotny.
\newblock Waveguide-integrated van der waals heterostructure photodetector at telecom wavelengths with high speed and high responsivity.
\newblock {\em Nat. Nanotechnol.}, 15(2):118--124, 2020.

\bibitem{marin:2019}
JF~Gonzalez Marin, Dmitrii Unuchek, Kenji Watanabe, Takashi Taniguchi, and Andras Kis.
\newblock Mos2 photodetectors integrated with photonic circuits.
\newblock {\em npj 2D Mater. Appl.}, 3(1):14, 2019.

\bibitem{ling:2023}
Haonan Ling, Arnab Manna, Jialiang Shen, Ho-Ting Tung, David Sharp, Johannes Fr{\"o}ch, Siyuan Dai, Arka Majumdar, and Artur~R Davoyan.
\newblock Deeply subwavelength integrated excitonic van der waals nanophotonics.
\newblock {\em Optica}, 10(10):1345--1352, 2023.

\bibitem{li:2021}
Chi Li, Johannes~E Froch, Milad Nonahal, Thinh~N Tran, Milos Toth, Sejeong Kim, and Igor Aharonovich.
\newblock Integration of hbn quantum emitters in monolithically fabricated waveguides.
\newblock {\em ACS Photonics}, 8(10):2966--2972, 2021.

\bibitem{dolado:2020}
Irene Dolado, Francisco~Javier Alfaro-Mozaz, Peining Li, Elizaveta Nikulina, Andrei Bylinkin, Song Liu, James~H Edgar, Felix Casanova, Luis~E Hueso, Pablo Alonso-Gonz{\'a}lez, et~al.
\newblock Nanoscale guiding of infrared light with hyperbolic volume and surface polaritons in van der waals material ribbons.
\newblock {\em Adv. Mater.}, 32(9):1906530, 2020.

\bibitem{kim:2023}
Minsu Kim, Kyung~Yeol Ma, Hyeongjoon Kim, Yeonju Lee, Jong~Hyun Park, and Hyeon~Suk Shin.
\newblock 2d materials in the display industry: status and prospects.
\newblock {\em Adv. Mater.}, 35(43):2205520, 2023.

\bibitem{li:2024}
Lu~Li, Qinqin Wang, Fanfan Wu, Qiaoling Xu, Jinpeng Tian, Zhiheng Huang, Qinghe Wang, Xuan Zhao, Qinghua Zhang, Qinkai Fan, et~al.
\newblock Epitaxy of wafer-scale single-crystal mos2 monolayer via buffer layer control.
\newblock {\em Nat. Commun.}, 15(1):1825, 2024.

\bibitem{najafidehaghani:2021}
Emad Najafidehaghani, Ziyang Gan, Antony George, Tibor Lehnert, Gia~Quyet Ngo, Christof Neumann, Tobias Bucher, Isabelle Staude, David Kaiser, Tobias Vogl, et~al.
\newblock 1d p--n junction electronic and optoelectronic devices from transition metal dichalcogenide lateral heterostructures grown by one-pot chemical vapor deposition synthesis.
\newblock {\em Adv. Funct. Mater.}, 31(27):2101086, 2021.

\bibitem{cai:2018}
Zhengyang Cai, Bilu Liu, Xiaolong Zou, and Hui-Ming Cheng.
\newblock Chemical vapor deposition growth and applications of two-dimensional materials and their heterostructures.
\newblock {\em Chem. Rev.}, 118(13):6091--6133, 2018.

\bibitem{masubuchi:2018}
Satoru Masubuchi, Masataka Morimoto, Sei Morikawa, Momoko Onodera, Yuta Asakawa, Kenji Watanabe, Takashi Taniguchi, and Tomoki Machida.
\newblock Autonomous robotic searching and assembly of two-dimensional crystals to build van der waals superlattices.
\newblock {\em Nat. Commun.}, 9(1):1413, 2018.

\bibitem{mannix:2022}
Andrew~J Mannix, Andrew Ye, Suk~Hyun Sung, Ariana Ray, Fauzia Mujid, Chibeom Park, Myungjae Lee, Jong-Hoon Kang, Robert Shreiner, Alexander~A High, et~al.
\newblock Robotic four-dimensional pixel assembly of van der waals solids.
\newblock {\em Nat. Nanotechnol.}, 17(4):361--366, 2022.

\bibitem{kong:2019}
Wei Kong, Hyun Kum, Sang-Hoon Bae, Jaewoo Shim, Hyunseok Kim, Lingping Kong, Yuan Meng, Kejia Wang, Chansoo Kim, and Jeehwan Kim.
\newblock Path towards graphene commercialization from lab to market.
\newblock {\em Nat. Nanotechnol.}, 14(10):927--938, 2019.

\bibitem{andryieuski:2012}
Andrei Andryieuski, Radu Malureanu, Giulio Biagi, Tobias~Holmgaard St{\ae}r, and Andrei Lavrinenko.
\newblock Compact dipole nanoantenna coupler to plasmonic slot waveguide.
\newblock {\em Opt. Lett.}, 37(6):1124--1126, 2012.

\bibitem{smith:2015}
Cameron~LC Smith, Nicolas Stenger, Anders Kristensen, N~Asger Mortensen, and Sergey~I Bozhevolnyi.
\newblock Gap and channeled plasmons in tapered grooves: a review.
\newblock {\em Nanoscale}, 7(21):9355--9386, 2015.

\bibitem{ayata:2017}
Masafumi Ayata, Yuriy Fedoryshyn, Wolfgang Heni, Benedikt Baeuerle, Arne Josten, Marco Zahner, Ueli Koch, Yannick Salamin, Claudia Hoessbacher, Christian Haffner, et~al.
\newblock High-speed plasmonic modulator in a single metal layer.
\newblock {\em Science}, 358(6363):630--632, 2017.

\bibitem{kumar:2020}
Shailesh Kumar, Till Lei{\ss}ner, Sergejs Boroviks, Sebastian~KH Andersen, Jacek Fiutowski, Horst-Gunter Rubahn, N~Asger Mortensen, and Sergey~I Bozhevolnyi.
\newblock Efficient coupling of single organic molecules to channel plasmon polaritons supported by v-grooves in monocrystalline gold.
\newblock {\em ACS Photonics}, 7(8):2211--2218, 2020.

\bibitem{thomaschewski:2021}
Martin Thomaschewski, Christian Wolff, and Sergey~I Bozhevolnyi.
\newblock High-speed plasmonic electro-optic beam deflectors.
\newblock {\em Nano Lett.}, 21(9):4051--4056, 2021.

\bibitem{yezekyan:2024}
Torgom Yezekyan, Martin Thomaschewski, Paul Conrad~Vaagen Thrane, and Sergey~I Bozhevolnyi.
\newblock Plasmonic electro-optic modulators on lead zirconate titanate platform.
\newblock {\em Nanophotonics}, (0), 2024.

\bibitem{kewes:2016}
G{\"u}nter Kewes, Max Schoengen, Oliver Neitzke, Pietro Lombardi, Rolf-Simon Sch{\"o}nfeld, Giacomo Mazzamuto, Andreas~W Schell, J{\"u}rgen Probst, Janik Wolters, Bernd L{\"o}chel, et~al.
\newblock A realistic fabrication and design concept for quantum gates based on single emitters integrated in plasmonic-dielectric waveguide structures.
\newblock {\em Sci. Rep.}, 6(1):28877, 2016.

\bibitem{kumar:2021}
Shailesh Kumar and Sergey~I Bozhevolnyi.
\newblock Single photon emitters coupled to plasmonic waveguides: A review.
\newblock {\em Advanced Quantum Technologies}, 4(10):2100057, 2021.

\bibitem{huang:2010}
Jer-Shing Huang, Victor Callegari, Peter Geisler, Christoph Br{\"u}ning, Johannes Kern, Jord~C Prangsma, Xiaofei Wu, Thorsten Feichtner, Johannes Ziegler, Pia Weinmann, et~al.
\newblock Atomically flat single-crystalline gold nanostructures for plasmonic nanocircuitry.
\newblock {\em Nat. Commun.}, 1(1):150, 2010.

\bibitem{chen:2010}
Yuntian Chen, Torben~Roland Nielsen, Niels Gregersen, Peter Lodahl, and Jesper M{\o}rk.
\newblock Finite-element modeling of spontaneous emission of a quantum emitter at nanoscale proximity to plasmonic waveguides.
\newblock {\em Physical Review B—Condensed Matter and Materials Physics}, 81(12):125431, 2010.

\bibitem{chen:2016}
Haitao Chen, Jiong Yang, Evgenia Rusak, Jakob Straubel, Rui Guo, Ye~Win Myint, Jiajie Pei, Manuel Decker, Isabelle Staude, Carsten Rockstuhl, et~al.
\newblock Manipulation of photoluminescence of two-dimensional mose2 by gold nanoantennas.
\newblock {\em Sci. Rep.}, 6(1):22296, 2016.

\bibitem{indukuri:2020}
SRK~Chaitanya Indukuri, Christian Frydendahl, Jonathan Bar-David, Noa Mazurski, and Uriel Levy.
\newblock Ws2 monolayers coupled to hyperbolic metamaterial nanoantennas: broad implications for light--matter-interaction applications.
\newblock {\em ACS Appl. Nano Mater.}, 3(10):10226--10233, 2020.

\bibitem{volkov:2009}
Valentyn~S Volkov, Jacek Gosciniak, Sergey~I Bozhevolnyi, SG~Rodrigo, L~Martin-Moreno, FJ~Garcia-Vidal, E~Devaux, and TW~Ebbesen.
\newblock Plasmonic candle: towards efficient nanofocusing with channel plasmon polaritons.
\newblock {\em New J. Phys.}, 11(11):113043, 2009.

\bibitem{johnson:1972}
PB~Johnson and RW~Christy.
\newblock Optical constants of the noble metals.
\newblock {\em Phys. Rev. B}, 6(12):4370, 1972.

\end{thebibliography}


\subsection{Funding:} C.F. was supported by the Carlsberg Foundation as an Reintegration Fellow (Grant No.: CF21-0216) during this work, while also acknowledging the VILLUM Foundation (Grant No. 58634). The Center for Polariton-driven Light–Matter Interactions (POLIMA) is sponsored by the Danish National Research Foundation (Project No. DNRF165).

\subsection{Author contributions:} C.F. and S.I.B conceived of the project. Waveguide fabrication was done by T.Y. 2D material transfer and exfoliation was performed by C.F. Optical measurements were done by T.Y. and C.F. Optical simulations and waveguide design was done by C.F. and V.A.Z. S.I.B supervised the project. All authors contributed to the interpretation of results and the writing of the manuscript.

\subsection{Conflicts of interests:} The authors declare no competing financial interests.

\subsection{Data availability:} All raw data are available upon reasonable request.

\subsection{Correspondence:} cfry@phys.au.dk and seib@mci.sdu.dk

\end{document}